%% file: jmlr.tex
\documentclass[pmlr,twocolumn,10pt]{jmlr} 
\usepackage{booktabs}
\usepackage{hyperref}
\usepackage{bm}
\usepackage{url}
\usepackage{xcolor}
\usepackage{bigstrut}
\usepackage{multicol}
\usepackage{stfloats}
\usepackage{multirow}
\usepackage{balance}

\DeclareMathOperator*{\argmin}{arg\,min}
\usepackage{algorithm}
\usepackage[noend]{algpseudocode}

\algnewcommand\algorithmicinput{\textbf{Input:}}
\algnewcommand\Input{\item[\algorithmicinput]}
\algnewcommand\algorithmicoutput{\textbf{Output:}}
\algnewcommand\Output{\item[\algorithmicoutput]}
\algnewcommand\algorithmicinitialize{\textbf{Initialize:}}
\algnewcommand\Initialize{\item[\algorithmicinitialize]}
\usepackage{float}
\usepackage{graphicx}
\usepackage{subfigure}
\usepackage{enumitem, enumerate}
\usepackage{siunitx}
\newcommand{\ours}{\textsc{PTGB}\xspace}

\theorembodyfont{\upshape}
\theoremheaderfont{\scshape}
\theorempostheader{:}
\theoremsep{\newline}

\jmlrvolume{LEAVE UNSET}
\jmlryear{2023}
\jmlrsubmitted{LEAVE UNSET}
\jmlrpublished{LEAVE UNSET}
\jmlrworkshop{Conference on Health, Inference, and Learning (CHIL) 2023}
\title[\ours: Pre-Train Graph Neural Networks for Brain Network Analysis]{\ours: Pre-Train Graph Neural Networks for \\ Brain Network Analysis}

\author{
\Name{Yi Yang} \Email{yi.yang@emory.edu}\\
\addr Emory University, USA
\AND
\Name{Hejie Cui}
\Email{hejie.cui@emory.edu}\\
\addr Emory Unversity, USA
\AND
\Name{Carl Yang} \Email{j.carlyang@emory.edu}\\
\addr Emory Unversity, USA
}

\begin{document}

\maketitle

\begin{abstract}
The human brain is the central hub of the neurobiological system, controlling behavior and cognition in complex ways. Recent advances in neuroscience and neuroimaging analysis have shown a growing interest in the interactions between brain regions of interest (ROIs) and their impact on neural development and disorder diagnosis. As a powerful deep model for analyzing graph-structured data, Graph Neural Networks (GNNs) have been applied for brain network analysis. However, training deep models requires large amounts of labeled data, which is often scarce in brain network datasets due to the complexities of data acquisition and sharing restrictions. 
To make the most out of available training data, we propose \ours, a GNN pre-training framework that captures intrinsic brain network structures, regardless of clinical outcomes, and is easily adaptable to various downstream tasks. \ours comprises two key components: (1) an unsupervised pre-training technique designed specifically for brain networks, which enables learning from large-scale datasets without task-specific labels; (2) a data-driven parcellation atlas mapping pipeline that facilitates knowledge transfer across datasets with different ROI systems.
Extensive evaluations using various GNN models have demonstrated the robust and superior performance of \ours compared to baseline methods.
\end{abstract}

\paragraph*{Data and Code Availability}
The empirical study in this work uses three real-world brain network datasets: 1) the Bipolar Disorder (BP) dataset, 2) the Human Immunodeficiency Virus Infection (HIV) dataset, and 3) the Parkinson's Progression Markers Initiative (PPMI) dataset. The BP and HIV are local datasets, while the large-scale PPMI dataset\footnote{\url{https://www.ppmi-info.org/}} is publicly available for authorized users. We followed the data preprocessing pipelines provided by the open-source BrainGB platform\footnote{\url{https://braingb.us/}} \citep{cui2022braingb} for the construction of brain networks based on raw neuroimaging data. The full implementation of this work is publicly available at \url{https://github.com/Owen-Yang-18/BrainNN-PreTrain}.

\paragraph*{Institutional Review Board (IRB)}
The study has been approved by an Institutional Review Board (IRB) to ensure the ethical and responsible use of human subjects in research. The IRB reviewed and approved the study protocols and consent forms, ensuring that the rights and welfare of the participants are protected. The study strictly adheres to the Good Clinical Practice guidelines and U.S. 21 CFR Part 50 (Protection of Human Subjects) to ensure the safety and privacy of the participants. All the data used in this work is processed anonymously to protect the privacy of participants, and no personally identifiable information is used or disclosed. 

\input{Sections/01_introduction.tex}
\input{Sections/02_related_work.tex}
\input{Sections/03_method.tex}
\input{Sections/04_experiment.tex}
\input{Sections/05_conclusions.tex}

\balance
\bibliography{jmlr}

\clearpage
\appendix
\input{Sections/06_appendix.tex}

\end{document}

%% file: Sections/01_introduction.tex
\section{Introduction}
Brain network analysis has attracted considerable interest in neuroscience studies in recent years. A brain network is essentially a connected graph constructed from different raw imaging modalities such as Diffusion Tensor Imaging (DTI) and functional Magnetic Resonance Imaging (fMRI), where nodes are composed by the anatomical regions of interest (ROIs) given predefined parcellation atlas, and connections are usually formed with the correlations among ROIs. Effective brain network analysis plays a pivotal role in understanding the biological structures and functions of complex neural systems, which potentially helps the early diagnosis of neurological disorders and facilitates neuroscience research \citep{maartensson2018stability, yahata2016small, lindquist2008statistical, smith2012future}. 

Graph Neural Networks (GNNs) have emerged as a powerful tool for analyzing graph-structured data, delivering impressive results on a wide range of network datasets, including social networks, recommender systems, knowledge graphs, protein and gene networks, and molecules, among others \citep{kipf2016semi, hamilton2017inductive, schlichtkrull2018modeling, vashishth2020composition, xu2019powerful, ying2018hierarchical, zhang2020gcn, liu2021pre, xiong2019pretrained, cui2022can, xu2022counterfactual}. These models have proven their ability to learn powerful representations and efficiently compute complex graph structures, making them well-suited for various downstream tasks. In the field of neuroscience, GNN has been applied to brain network analysis, specifically for graph-level classification/regression \citep{ying2018hierarchical, xu2019powerful, errica2020fair, luo2022multi, dai2022transformer,xu2023neighborhood} and important vertex/edge identification \citep{ying2019gnnexplainer, luo2020parameterized, vu2020pgm, yu2022learning, kan2022bracenet}, towards tasks such as connectome-based disease prediction and multi-level neural pattern discovery. 
However, deep learning models, including GNNs, require large amounts of labeled data to achieve optimal performance \citep{hu2020strategies, you2020graph, zhu2021transfer}. While neuroimaging datasets are available from national neuroimaging studies such as the ABCD \citep{casey2018adolescent}, ADNI \citep{hinrichs2009spatially}, and PPMI \citep{aleksovski2018disease}, these datasets are still relatively small compared to graph datasets from other domains, such as datasets with 41K to 452K graphs on OGB \citep{hu2020open} and datasets with thousands to millions of graphs on NetRepo \citep{rossi2016interactive}). The limited amount of data can result in overfitting when training deep models.

Transfer learning offers a solution to the challenge of limited data availability in training deep models. It allows a model pre-trained on large-scale source datasets to be adapted to smaller target datasets while maintaining robust performance. However, the success of transfer learning depends on the availability of similar supervision labels on the source and target dataset. This is not always feasible in large-scale public studies, particularly in the field of brain network analysis. Self-supervised pre-training has been shown to be effective in various domains, such as computer vision \citep{he2020momentum, chen2020simple}, natural language processing \citep{devlin2019bert, yu2022coco}, and graph mining \citep{10.1145/3534678.3539249}. We aim to explore a self-supervised pre-training approach for GNNs on brain networks that is not restricted by task-specific supervision labels. Despite the promising potential, unique challenges still need to be addressed to achieve effective disease prediction. One of the major challenges is the inconsistent ROI parcellation systems in constructing different brain network datasets, which hinders the transferability of pre-trained models across datasets. 
The process of parcellating raw imaging data into brain networks is highly complex and usually done ad hoc by domain experts for each study, making it unrealistic to expect every institution to follow the same parcellation system. Although some institutions may release preconstructed brain network datasets \citep{di2014autism}, the requirement for universal adherence to a single parcellation system is infeasible.

To tackle the challenge of insufficient training data for GNNs in brain network analysis, we present \textbf{P}re-\textbf{T}raining \textbf{G}raph neural networks for \textbf{B}rain networks (PTGB), a fully unsupervised pre-training approach that captures shared structures across brain network datasets. \ours adapts the data-efficient MAML \citep{finn2017model} with a two-level contrastive learning strategy based on the naturally aligned node systems of brain networks across individuals.
Additionally, to overcome the issue of diverse parcellation systems, we introduce a novel data-driven atlas mapping technique. This technique transforms the original features into low-dimensional representations in a uniform embedding space and aligns them using variance-based projection, which incorporates regularizations that preserve spatial relationships, consider neural modules, and promote sparsity.

In summary, our contributions are three-folded: 
\begin{itemize}[nosep,leftmargin=*]
\item We present an unsupervised pre-training approach for GNNs on brain networks, addressing the issue of resource-limited training.
\item We propose a two-level contrastive sampling strategy tailored for GNN pre-training on brain networks, which combines with a data-driven brain atlas mapping strategy that employs customized regularizations and variance-based sorting to enhance cross-dataset learning. 
\item Our experiments against shallow and deep baselines demonstrate the effectiveness of our proposed \ours. Further, we provide an in-depth analysis to understand the influence of each component.
\end{itemize}

%% file: Sections/02_related_work.tex
\section{Related Work}
\paragraph*{GNNs for Brain Network Analysis.}
GNNs are highly effective for analyzing graph-structured data and there have been some pioneering attempts to use them for predicting diseases by learning over brain networks. For example, BrainGNN \citep{Li:2021fa} proposes ROI-aware graph convolutional layers and ROI-selection pooling layers for predicting neurological biomarkers.
BrainNetCNN \citep{kawahara2017brainnetcnn} designs a CNN that includes edge-to-edge, edge-to-node, and node-to-graph convolutional filters, leveraging the topological locality of brain connectome structures.
BrainNetTF \citep{kan2022bnt} introduces a transformer architecture with an orthonormal clustering readout function that considers ROI similarity within functional modules.
Additionally, various studies \citep{cui2022interpretable, kan2022fbnetgen, zhu2022joint, cui2022braingb,yu2022learning} have shown that, when data is sufficient, GNNs can greatly improve performance in tasks such as disease prediction. 
However, in reality, the lack of training data is a common issue in neuroscience research, particularly for specific domains and clinical tasks~\citep{xu2023weakly}. 
Despite this, there has been little research into the ability of GNNs to effectively train for brain network analysis when data is limited. 

\paragraph*{Unsupervised Graph Representation Learning and GNN Pre-training.}
\label{sec: unsupervised related work}
Unsupervised learning is a widely used technique for training complex models when resources are limited. Recent advancements in contrastive learning \citep{10.5555/3524938.3525087, he2020momentum, yu2021fine, zhu2022structure} have led to various techniques for graphs.
For instance, GBT \citep{bielak2022graph} designs a Barlow Twins \cite{zbontar2021barlow} loss function based on the empirical cross-correlation of node representations learned from two different views of the graph \citep{zhao2021data}. 
Similarly, GraphCL \citep{you2020graph} involves a comparison of graph-level representations obtained from two different augmentations of the same graph.
DGI \citep{velickovic2019deep} contrasts graph and node representations learned from the original graph and its corruption.

To obtain strong models for particular downstream tasks, unsupervised training techniques can be used to pre-train a model, which is then fined tuned on the downstream tasks to reduce the dependence on labeled training data. 
The approach has proven highly successful in computer vision \citep{cao2020parametric, grill2020bootstrap},  natural language processing \citep{devlin2019bert, radford2018improving, radford2021learning, liang2020bond}, and  multi-modality (e.g. text-image pair) learning \citep{li2022blip, yao2021filip}.
There are various strategies for pre-training GNNs as well.
GPT-GNN \citep{hu2020gpt} proposes graph-oriented pretext tasks, such as masked attribute and edge reconstruction.
L2P-GNN \citep{lu2021learning} introduces dual adaptation by simultaneously optimizing the encoder on a node-level link prediction objective and a graph-level self-supervision task similar to DGI.
Others, such as GMPT \citep{hou2022neural} adopt an inter-graph message-passing approach to obtain context-aware node embedding and optimize the model concurrently under supervision and self-supervision.
To the best of our knowledge, the effectiveness of both contrastive learning and pre-training has not been investigated in the context of the unique properties of brain networks.

%% file: Sections/03_method.tex
\section{Unsupervised Brain Network Pre-training}
\textbf{Problem Definition.}
The available training resource includes a collection of brain network datasets $\mathcal{S} = \{ \mathcal{D}_1, \mathcal{D}_2, \cdots \mathcal{D}_s\}$, where each dataset contains a varying number of brain networks. We consider each brain network instance with $M$ number of defined ROIs as an undirected weighted graph $\mathcal{G}$ with $M$ nodes. $\mathcal{G}$ is represented by a node-set $\mathcal{V} = \{v_m\}_{m=1}^{M}$, an edge set $\mathcal{E} = \mathcal{V} \times \mathcal{V}$, and a weighted adjacency matrix $\bm{A} \in \mathbb{R}^{M\times M}$. 
We define a $\theta$ parameterized GNN model $f(\cdot)$, 
and our goal is to propose a pre-training schema that can effectively learn an initialization $\theta_0$ for $f(\cdot)$ on a set of source datasets $\mathcal{S}_{\text{source}} \subset \mathcal{S}$ via self-supervision and adapt $f_{\theta_0}(\cdot)$ to a local optimum $\theta^*$ on a target set $\mathcal{S}_{\text{target}} \in \mathcal{S}$.

\begin{figure*}
    \centering
    \includegraphics[width=\textwidth]{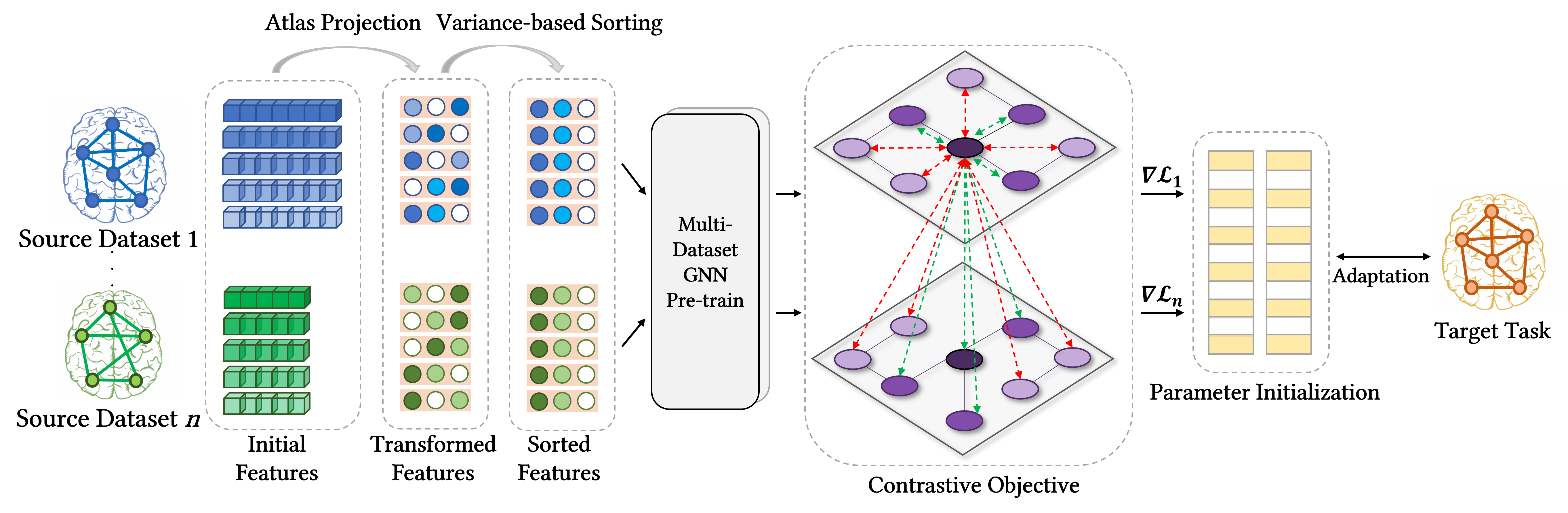}
    \vspace*{-5pt}
    \caption{Overview of the proposed framework \ours. The initial features of the source datasets are projected to a fixed dimension through atlas transformation followed by variance-based feature alignment, which facilitates self-supervised GNN pre-training on multiple datasets via the novel two-level contrastive learning objective. The learned model can serve as the parameter initialization and be further fine-tuned on target tasks.}
    \label{fig:pipeline}
    \vspace*{-10pt}
\end{figure*}%

\subsection{GNN Pre-training for Brain Networks}
\label{sec:multi}
The goal of pre-training a GNN model for brain networks is to learn an appropriate initialization that can easily be adapted to downstream task.
Note that the concept of pre-training is distinct from transfer learning since the latter expects a similarity between the source and target data as well as their learning objectives (\emph{e.g.,} loss functions), while this is often lacking in brain network analysis due to absence of sufficient ground truth labels in large scale studies as well as inherent differences in their brain network parcellation methods across varying datasets.
Practically, a GNN model can be pre-trained either on a singular task with a single source dataset or on a collection of tasks with multiple source datasets.
The proposed \ours framework adopts the latter option since multi-task pre-training reduces the likelihood of the model being biased towards the knowledge of data from a singular source, which could be particularly concerning if the source and target data shares limited similarity leading to poor downstream adaptation due to information loss during model transfer. 
However, a naive approach towards multi-task pre-training would not suffice in learning a robust model initialization.
Specifically, it presents two underlying risks: (1) the model may not perform consistently well on all tasks and may also overfit to a particular task which significantly undermines model generalizability; and (2) the process could be computationally inefficient with increasing number of tasks regardless if the model is optimized sequentially or simultaneously on all tasks \citep{yang2022data}. 

To this end, we adopt the popular data-efficient training techniques presented in MAML \citep{finn2017model} with the goal of ensuring consistent performance on all tasks as well as computational efficiency.
The MAML technique is characterized by an inner-loop adaptation and an outer-loop update \citep{raghu2019rapid}. 
At each training iteration, each input dataset is partitioned into an inner-loop support set and an outer-loop query set. The model is first trained on the support set without explicitly updating the parameters. Instead, the updates are temporarily stored as fast weights \citep{ba2016using}. These fast weights are then used to evaluate the query set and compute the actual gradients.
This approach makes use of approximating higher-order derivatives \citep{Tan_2019} at each step, allowing the model to foresee its optimization trajectory a few steps ahead, which practically reduces the number of required training iterations to reach local optima.
In our scenario, the joint optimization involves summing the loss over each brain network dataset, i.e., for $n$ number of datasets and their respective temporary fast weights $\{\theta'_i\}_{i=1}^n$ and outer-loop queries $\{\text{query}_i\}_{i=1}^n$, the step-wise update of the model parameter at time $t$ is 
$
\theta^{t+1} = \theta^t - \alpha \nabla_{\theta^t} \sum_{i=1}^n \mathcal{L}_{\text{query}_i} f_{\theta^{'t}_i}(\cdot). 
$
We hereby summarize this process in Algorithm \ref{alg:maml}.
In addition, we will also demonstrate the advantages of MAML-styled pre-training over vanilla multi-task pre-training as well as single task pre-training through experiments which will be discussed in Section \ref{sec:RQ1}.

\begin{algorithm}
\caption{MAML-based Multi-task Pre-training}\label{alg:maml}
\begin{algorithmic}[1]
\Input \text{Source task pool $S_\tau$, GNN model $f_{\theta}(\cdot)$}
\Require \text{$\alpha$, $\beta$: learning rate hyperparameters}
\State \text{Randomly initialize $\theta$}\\
\While{not done}{
\For{each task $\tau_i$ in $S_\tau$}{

Sample a set of $k$ datapoints $\mathcal{D}_i$ from $\tau_i$ as support set

Evaluate the gradient for the task-wise objective $\nabla_\theta \mathcal{L}_{\mathcal{D}_i}f(\theta)$

Compute the inner-loop adapted parameters $\theta'_i \gets \theta - \beta \nabla_\theta \mathcal{L}_{\mathcal{D}_i}f(\theta)$

Sample another set of datapoints $\mathcal{D}'_i$ from $\tau_i$ as query set

}

Update the GNN model parameters $\theta \gets \theta - \alpha \nabla_\theta \sum_{\mathcal{D}'_i \sim S_\tau} \mathcal{L}_{\mathcal{D}'_i}f_{\theta'_i}(\cdot)$

}
\end{algorithmic}
\end{algorithm}

\subsection{Brain Network Oriented Two-Level Contrastive Learning}
\label{sec:twolevel}
\begin{figure}
    \centering
    \includegraphics[width=0.85\linewidth]{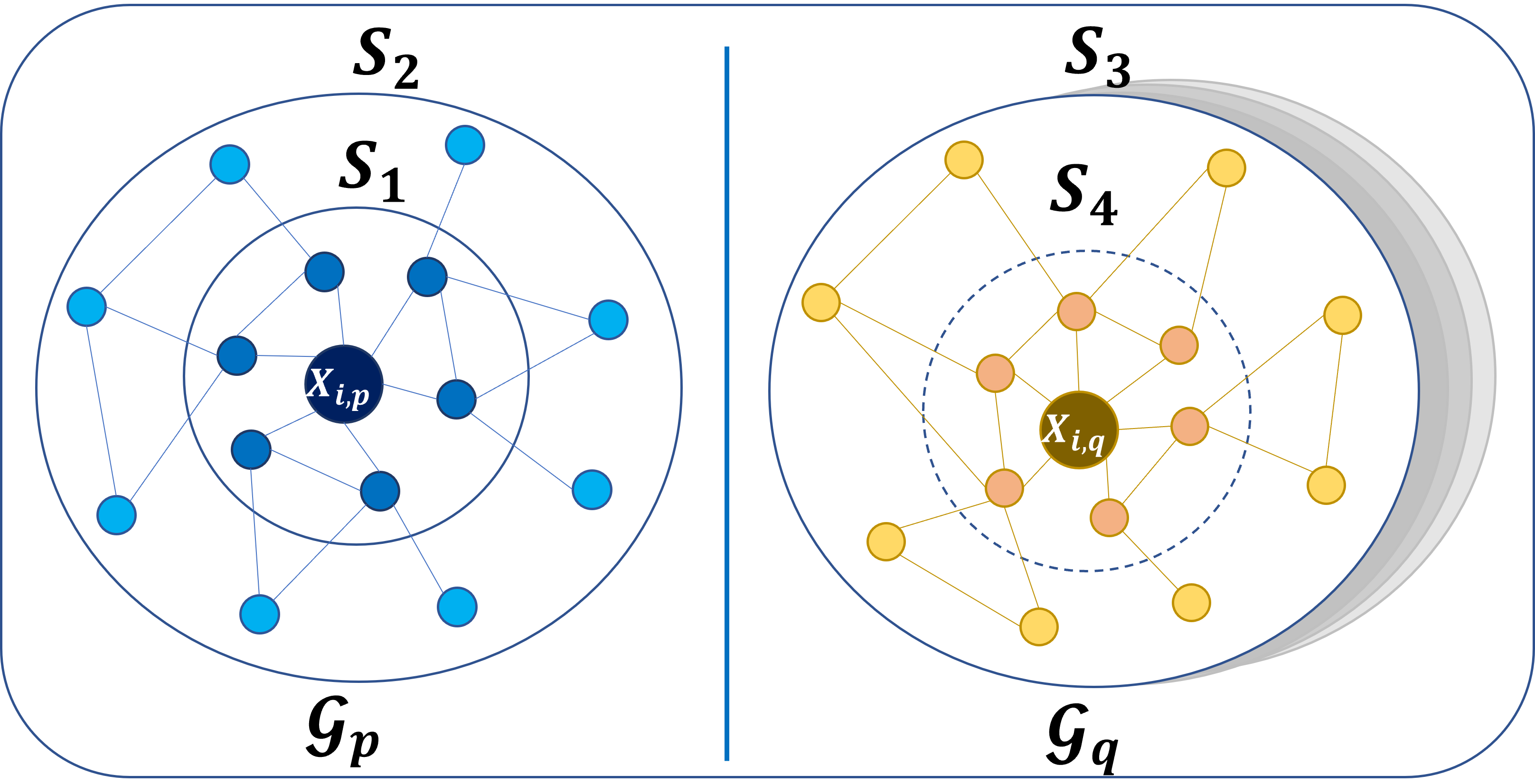}
    \caption{Visual demonstration of the sample types where $X_{i,p}$ is the anchor and $\mathbf{S_1}$/$\mathbf{S_4}$ are sampled as 1-hop neighbors.}
    \label{fig:samples}
    \vspace*{-10pt}
\end{figure}

Given the high cost of acquiring labeled training data for brain network analysis, our pre-training pipeline of \ours adopts to the effective label-free learning strategy of contrastive learning (CL).
CL aims to maximize the mutual information (MI) between an anchor point of investigation $X$ from a data distribution $\mathcal{H}$ and its positive samples $X^+$, while minimizing MI with its negative samples $X^-$. The contrastive objective function is formulated as follows:
\begin{equation}
    \mathcal{J}_\text{con} = \argmin \left[ \left( -I(X; X^+) + I(X; X^-) \right) \right].
\end{equation}%
In the context of graph CL, given an anchor node representation $z_\alpha$, a set of positive samples $\mathbf{S}^+$, and a set of negative samples $\mathbf{S}^-$, the training objective is based on the Jensen-Shannon divergence \citep{hjelm2018learning}, 
\begin{equation}
     \mathcal{J}_{\text{JSD}}(z_\alpha) = \argmin \left[\left( -I(z_\alpha; \mathbf{S}^+) + I(z_\alpha; \mathbf{S}^-)\right)\right],
\end{equation}
where 
\begin{align}
    I(z_\alpha; \mathbf{S}^+) &=\frac{1}{|\mathbf{S}^+|}\sum_{z_{s^+} \in \mathbf{S}^+}\text{sp} \left( \frac{z_\alpha^\top z_{s^+}}{\lVert z_\alpha \rVert \lVert z_{s^+} \rVert}\right),\\
    I(z_\alpha; \mathbf{S}^-) &= \frac{1}{|\mathbf{S}^-|}\sum_{z_{s^-} \in \mathbf{S}^-}\text{sp} \left( \frac{z_\alpha^\top z_{s^-}}{\lVert z_\alpha \rVert \lVert z_{s^-} \rVert}\right),
\end{align}\\
and $\text{sp}(\cdot) = \log (1 + e^{\cdot})$ is softplus nonlinearity.\\
\hfill \break
The ultimate goal of our framework is to localize effective GNN CL learning \citep{Zhu:2021wh} for brain networks. 
Given a dataset $\mathcal{D}$ and an anchor node $i$ from graph $\mathcal{G}_p \in \mathcal{D}$ with the learned representation $z_{i, p}$, we propose to categorize the possible sample selections into three fundamental types (a visualization is shown in Figure \ref{fig:samples}):
\begin{itemize}[nosep,leftmargin=*]
    \item \underline{$\mathbf{S_1}$}: $\{z_{j, p}\,:\, j \in \mathcal{N}_k(i, p)\}$ refers to the node representation set within the the $k$-hop neighborhood of the anchor in graph $\mathcal{G}_p$.
    \item \underline{$\mathbf{S_2}$}: $\{z_{j, p}\,:\, j \notin \mathcal{N}_k(i, p) \}$ refers to the remaining node representation set in graph $\mathcal{G}_p$ that are not in the the $k$-hop neighborhood of the anchor.
    \item \underline{$\mathbf{S_3}$}: $\{z_{j, q}\,:\, \mathcal{G}_q \in \mathcal{D}, \, j \in \mathcal{G}_q, \, q \neq p \}$ refers to the node representation set of nodes in all the other graphs of dataset $\mathcal{D}$.
\end{itemize}

Notice that our framework leverages the $k$-hop substructure around the anchor node to further differentiate $\mathbf{S_1}$ and $\mathbf{S_2}$ for contrastive optimization. This design is driven by two considerations:
\textbf{(1) Regarding GNN learning.} Given that node representations are learned from the information aggregation of its $k$-hop neighborhood, maximizing the MI of an anchor to its $k$-hop neighbors naturally enhances lossless message passing of GNN convolutions.
\textbf{(2) Regarding the uniqueness of brain networks.} Brain networks can be anatomically segmented into smaller neural system modules \citep{cui2021brainnnexplainer}, thus capturing subgraph-level knowledge can provide valuable signals for brain-related analysis.

Building on these three fundamental types of samples, we take advantage of the property of brain networks that ROI identities and orders are fixed across samples to introduce an additional sample type. 
This encourages the GNN to extract shared substructure knowledge by evaluating the MI of an anchor against its presence in other graphs. 
Given an anchor representation $z_{i,p}$ of node $i$ from graph $\mathcal{G}_p \in \mathcal{D}$, the novel inter-graph sample type is defined as:
\begin{itemize}[nosep,leftmargin=*]
    \item \underline{$\mathbf{S_4}$}:$\{ z_{j,q}\,:\, j \in \mathcal{N}_k(i, q) \cap \mathcal{N}_k(i, p), \, \mathcal{G}_q \in \mathcal{D}, \, q \neq p \}$, refers to the node representation set within the $k$-hop neighborhood of node $i$ in all other graphs in $\mathcal{D}$. Conceptually, $\mathbf{S_4}$ is a special subset of $\mathbf{S_3}$.%
\end{itemize}
It is important to note that for an anchor node $i$, its $k$-hop neighborhood structures might not be identical among different graphs.
As a result, we only consider shared neighborhoods when evaluating the mutual information across multiple graphs.
\begin{figure}
    \centering
    \includegraphics[width=0.5\linewidth]{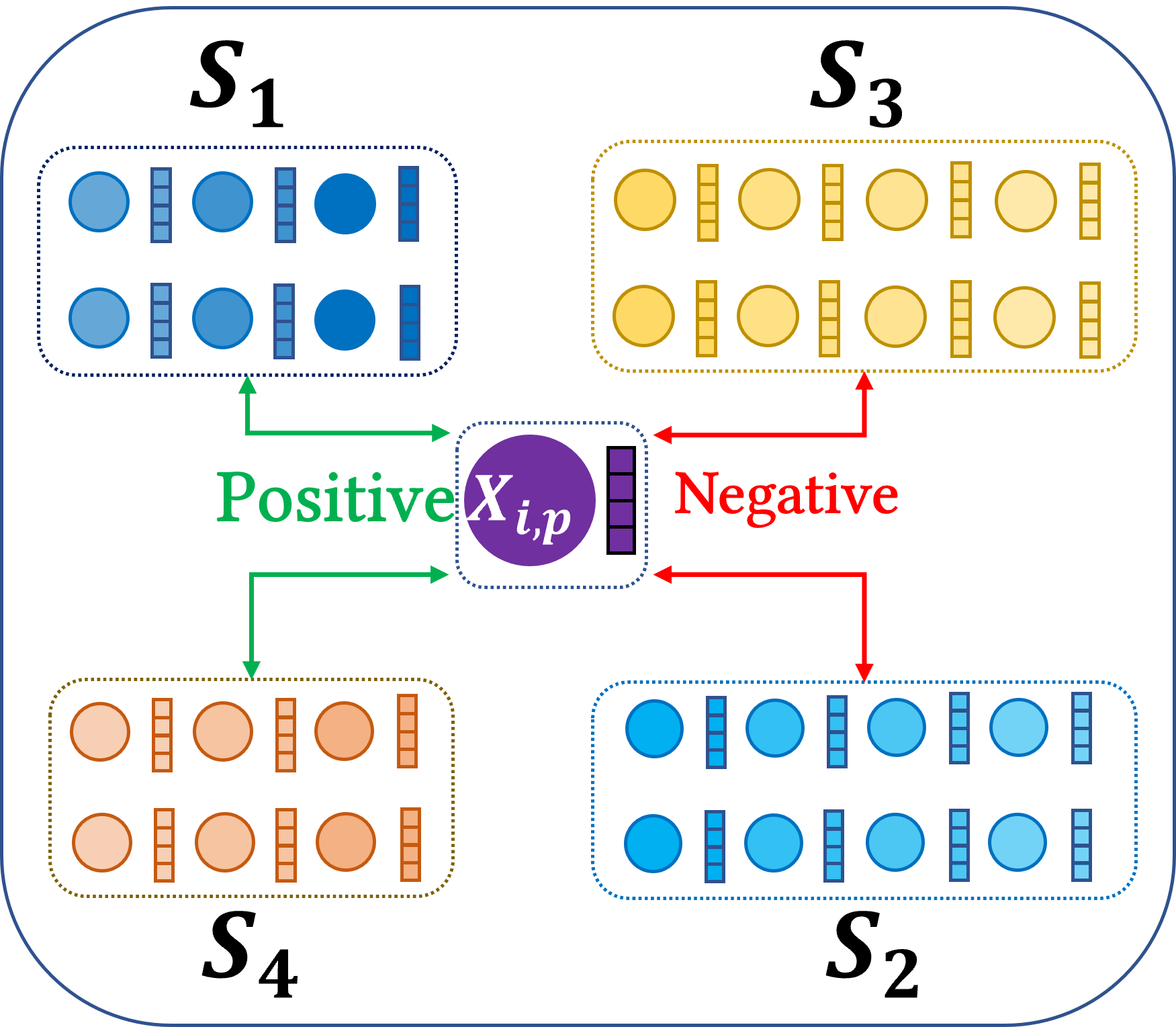}
    \caption{The sampling configuration of the proposed \ours framework. $\mathbf{S_1}$ and $\mathbf{S_4}$ are positive samples, $\mathbf{S_2}$ and the set $\mathbf{S_3} - \mathbf{S_4}$ are negative samples.}
    \label{fig:contrast}
    \vspace*{-15pt}
\end{figure}
To encourage the learning of unique neighborhood knowledge within a single brain network instance and shared substructure knowledge across the entire dataset, we configure $\mathbf{S_1}$ and $\mathbf{S_4}$ as positive samples while $\mathbf{S_2}$ and the set $\mathbf{S_3} - \mathbf{S_4}$ as negative samples, as illustrated in Figure \ref{fig:contrast}. Strictly speaking, $\mathbf{S_1}$ does not include the anchor itself, but the anchor is always a positive sample to itself by default.
\begin{table}
  \centering
  \caption{The sampling configuration of some existing graph contrastive learning methods. ``+'' denotes positive sampling, ``-'' for negative, and ``/'' for no consideration.}
   \vspace*{-5pt}
   \resizebox{0.5\linewidth}{!}{
    \begin{tabular}{|l|l|l|l|l|l|}
    \hline
   & $\mathbf{S_1}$  & $\mathbf{S_2}$  & $\mathbf{S_3}$  & $\mathbf{S_4}$ \bigstrut\\
    \hline
    DGI     & +     & +     & /     & / \bigstrut\\
    \hline
    InfoG     & +     & +     & --     & / \bigstrut\\
    \hline
    GCC        & +     & --     & --     & / \bigstrut\\
    \hline
    EGI       & +     & --     & --     & / \bigstrut\\
    \hline
    Ours     & +     & --     & --     & + \bigstrut\\
    \hline
    \end{tabular}
    }
  \label{tab:sample_summary}
  \vspace*{-10pt}
\end{table}
Furthermore, our sampling categorization can also help understand the objective formulations in various state-of-the-art graph CL frameworks \citep{velickovic2019deep, qiu2020gcc, xia2022progcl, sun2019infograph, zhu2021transfer}.
We summarize our findings in Table \ref{tab:sample_summary}.
Specifically, ``+''\, denotes positive sampling; ``-''\, denotes negative sampling; and ``/'' means that the sample type is not considered.
It can be observed that DGI and InfoGraph (InfoG) use graph representation pooled from node representations as a special sample, which is essentially equivalent to jointly considering $\mathbf{S_1}$ and $\mathbf{S_2}$ without explicit differentiation.
On the other hand, GCC and EGI, which are more closely related to our framework, leverage neighborhood mutual information maximization on a single graph, but fail to extend this to a multi-graph setting like ours.

\subsection{Data-driven Brain Atlas Mapping}
\label{sec:regularizers}
\paragraph{Motivation.}
When fine-tuning a pre-trained model on a new data domain, the misalignment between source and target signals can negatively impact its adaptation.
This issue is particularly relevant in brain networks, where it is hard, if not impossible, to require every brain network data provider to stick to the same brain atlas template, and each template can use a unique system of ROIs.
For instance, the HIV dataset we obtained is parcellated from the AAL90 template \citep{tzourio2002automated}, leading to 90 defined ROIs; while the PPMI dataset uses the Desikan-Killiany84 template \citep{desikan2006automated}, resulting in 84 defined ROIs.
As a result, brain networks in the two datasets will have different ROI semantics and graph structures. Although GNNs can handle graphs without fixed numbers and orders of nodes, constructing the most informative ROI (\emph{i.e.,} node) features as the connection profiles (\emph{i.e.,} adjacency) \citep{cui2022braingb, cui2022positional} can result in different feature dimensions and physical meanings. 
While manual conversion can be performed to translate between templates, it is a costly process that requires domain expertise to perform even coarse cross-atlas mappings.

To address this issue, we aim to provide a data-driven atlas mapping solution that is easily accessible and eliminates the strong dependency on network construction. The data-driven atlas mapping solution, which transforms the original node features into lower-dimensional representations that preserve the original connectivity information and align features across datasets, is learned independently on each dataset prior to GNN pre-training.

\subsubsection{Autoencoder with Brain Network Oriented Regularizers}
\ours adopts a one-layer linear autoencoder (AE)
as the base structure. The AE consists of a linear projection encoder $\mathbf{W}$ and a transposed decoder $\mathbf{W}^\top$, with the goal of learning a low-dimensional projection that can easily reconstruct the original presentation.
The loss function is defined as minimizing the reconstruction error $\mathcal{L}_{\text{rec}} = (1/M)\lVert \mathbf{X} - \mathbf{XW}\mathbf{W}^\top \rVert_2^2$, where $\mathbf{X} \in \mathbb{R}^{M \times M}$ is the input and $\mathbf{W} \in \mathbb{R}^{M \times D}$ is the learnable projection \citep{hinton1993autoencoders}.
To further enhance the feature compression and to guide the overall AE optimization, we propose to incorporate several regularizers that take into account the unique characteristics of brain networks.: 

\paragraph*{Locality-Preserving Regularizer (LR).}
We aim to ensure that the compressed features preserve the spatial relationships of the original brain surface.
To achieve this, we incorporate a locality preserving regularizer \citep{he2005neighborhood} to the AE objective. The regularizer is formulated as $\mathcal{L}_{\text{loc}} = (1/M) \lVert \mathbf{Y} - \mathbf{TY} \rVert^2$, where $\mathbf{Y} \in \mathbb{R}^{M \times D}$ represents the projected features from the AE and $\mathbf{T} \in \mathbb{R}^{M\times M}$ is a transition matrix constructed from the $k$-NN graph of the 3D coordinates of ROIs.

\paragraph*{Modularity-Aware Regularizer (CR).} 
Brain networks can be segmented into various neural system modules that characterize functional subsets of ROIs. In graph terminology, they are community structures. The projected feature should also capture information about neural system membership. However, obtaining ground-truth segmentations is a difficult task that requires expert knowledge. To overcome this challenge, we resort to community detection methods on graphs, specifically based on modularity maximization. 
The regularizer \citep{SALHAGALVAN2022474} is defined as minimizing
\begin{equation}
    \mathcal{L}_{\text{com}} = -\frac{1}{2D} \sum_{i,j=1}^M \left[ \mathbf{A}_{ij} - \frac{k_i k_j}{2D} \right] \exp(- \lVert y_i - y_j \rVert^2_2),
\end{equation} 
where $\mathbf{A} \in \mathbb{R}^{M\times M}$ is the graph adjacency matrix, $k_i$ denotes degree of node $i$, and $y_i$ is the AE projected features. Essentially, this optimization minimizes the $L_2$ distance between representations of nodes within the same communities, as measured by the modularity score, and maximizes the distance between representations of nodes in different communities.

\paragraph*{Sparsity-Oriented Regularizer (SC).}
Sparse networks have proven to be effective in learning robust representations from noisy data \citep{jeong2017dictionary, shi2019boosting, makhzani2013k}. In brain connectome analysis, sparsity has also been shown to improve the interpretation of task-specific ROI connections in generation and classification tasks \citep{kan2022fbnetgen}. 
To this end, we implement the popular KL-divergence smoothing to enforce sparsity in the parameters of the linear projection encoder, $\mathbf{W}$). This is formulated as: 
\begin{equation}
    \mathcal{L}_{\text{KL}} = \sum_{i=1}^M \sum_{j=1}^D \left[ \rho \log \left( \frac{\rho}{\hat{\rho}_{ij}} \right) + (1 - \rho) \log \left( \frac{1 - \rho}{1 - \hat{\rho}_{ij}} \right) \right],
\end{equation}
where $\rho$ is a small positive float set as the target sparsity value, and $\hat{\rho}_{ij}$ represents the element-wise activation of the encoder projection matrix $\mathbf{W} \in \mathbb{R}^{M\times D}$.

\subsubsection{Variance-based Dimension Sorting}
In addition to transforming dataset-specific features, cross-dataset alignment of feature signals is also crucial for improving model adaptation. The one-layer AE transforms the original feature vectors into weighted combinations of multiple dimensions, creating new feature dimensions which we name as \textit{virtual ROIs}. In the context of brain networks, this process helps to group ROIs and their signals. 
This idea is inspired by the well-studied functional brain modules \citep{philipson2002functional, anderson2004integrated, hilger2020temporal, brodmann1909vergleichende, zhou2020toolbox}, which provide a higher-level and generic organization of the brain surface, as opposed to fine-grained ROI systems.
Since the variations in ROI parcellations are due to differences in clinical conventions, it is reasonable to assume that there exists a shared virtual ROI system underlying different parcellation systems, similar to the discretization of functional brain modules. The community learning and neighborhood preserving regularizers, introduced in Section \ref{sec:regularizers}, allow us to capture these shared virtual ROIs in a data-driven manner.
Our ultimate goal is to align the discovered virtual ROIs across datasets, so that each virtual ROI characterizes the same functional module in the human brain, regardless of its origin. This cross-dataset alignment of virtual ROIs ensures that the model can effectively adapt to new datasets and provide meaningful insights into the different downstream analyses.

The objective of the one-layer linear AE is similar to PCA, as discussed in more detail in Appendix \ref{sec:pcaequiv}, with the added benefit of incorporating additional regularizers. PCA orders dimensions based on decreasing levels of sample variance \citep{hotelling1933analysis}. \ours leverage this approach by utilizing the learned parameters of the AE projection to estimate the variance of each virtual ROI (\textit{i.e.}, projected feature dimension). The sample variance of each virtual ROI indicates its representativeness of the original data variations. Given the shared patterns across different parcellation systems, we expect that similar virtual ROIs in datasets with different atlas templates will have similar variance scores, especially in terms of their order. By sorting the same number of virtual ROIs based on their sample variance in each dataset, we aim to align virtual ROI cross datasets, so that each virtual ROI represents the same functional unit in the human brain. The procedure is explained in detail in Algorithm \ref{alg:variance} in Appendix \ref{sec:varsorting}.

%% file: Sections/04_experiment.tex
\section{Experiments}
\begin{table*}[h!]
  \centering
  \caption{Disease prediction performance comparison. All results are averaged from 5-fold cross-validation along with standard deviations. The best result is highlighted in bold and runner-up is underlined. * denotes a significant improvement according to paired $t$-test with $\alpha$ = 0.05 compared with baselines.}
  \resizebox{0.99\textwidth}{!}{
    \begin{tabular}{llr|llllllll}
    \hline \bigstrut
    \multirow{2}[3]{*}{Type} & \multicolumn{2}{l}{\multirow{2}[3]{*}{Method}} & \multicolumn{2}{c}{BP-fMRI} & \multicolumn{2}{c}{BP-DTI} & \multicolumn{2}{c}{HIV-fMRI} & \multicolumn{2}{c}{HIV-DTI} \bigstrut[b]\\
\cline{4-11}              & \multicolumn{2}{l}{}  & \multicolumn{1}{c}{ACC} & \multicolumn{1}{c}{AUC} & \multicolumn{1}{c}{ACC} & \multicolumn{1}{c}{AUC} & \multicolumn{1}{c}{ACC} & \multicolumn{1}{c}{AUC} & \multicolumn{1}{c}{ACC} & \multicolumn{1}{c}{AUC} \bigstrut\\
    \hline
    NPT       & \multicolumn{2}{l|}{GCN} & 50.07\tiny{±13.70} & 50.11\tiny{±15.48} & 49.51\tiny{±14.68} & 51.83\tiny{±13.98} & 56.27\tiny{±15.84} & 57.16\tiny{±15.14} & 51.30\tiny{±16.42} & 53.82\tiny{±14.94} \bigstrut\\
    \hline
    \multirow{3}[2]{*}{NCL} & \multicolumn{2}{l|}{Node2Vec} & 48.51\tiny{±10.39} & 49.68\tiny{±7.23} & 50.83\tiny{±8.14} & 46.70\tiny{±10.33} & 52.61\tiny{±10.38} & 50.75\tiny{±10.94} & 49.65\tiny{±10.30} & 51.22\tiny{±10.79} \bigstrut[t]\\
              & \multicolumn{2}{l|}{DeepWalk} & 50.28\tiny{±9.33} & 51.59\tiny{±9.06} & 52.17\tiny{±9.74} & 48.36\tiny{±9.37} & 54.81\tiny{±11.26} & 55.55\tiny{±11.93} & 52.67\tiny{±11.42} & 50.88\tiny{±10.53} \\
              & \multicolumn{2}{l|}{VGAE} & 56.71\tiny{±9.68} & 55.24\tiny{±11.48} & 54.63\tiny{±12.09} & 54.21\tiny{±11.94} & 62.76\tiny{±9.47} & 61.25\tiny{±11.61} & 56.90\tiny{±9.72} & 55.35\tiny{±9.04} \bigstrut[b]\\
    \hline
    \multirow{3}[2]{*}{SCL} & \multicolumn{2}{l|}{GBT} & 57.21\tiny{±10.68} & 57.32\tiny{±10.48} & 56.29\tiny{±9.35} & 55.27\tiny{±10.54} & 65.73\tiny{±10.93} & 66.08\tiny{±10.43} & 59.80\tiny{±9.76} & 57.37\tiny{±9.49} \bigstrut[t]\\
              & \multicolumn{2}{l|}{GraphCL} & 59.79\tiny{±9.36} & 59.10\tiny{±10.78} & 57.57\tiny{±10.63} & 57.35\tiny{±9.67} & 67.08\tiny{±9.70} & 69.17\tiny{±10.68} & 60.43\tiny{±8.39} & 60.03\tiny{±10.48} \\
              & \multicolumn{2}{l|}{ProGCL} & 62.36\tiny{±8.90} & 62.61\tiny{±9.34} & 61.26\tiny{±8.37} & \underline{62.67\tiny{±8.46}} & 71.52\tiny{±9.19} & 72.16\tiny{±9.85} & \underline{62.48\tiny{±10.38}} & 61.94\tiny{±10.57} \bigstrut[b]\\
    \hline
    \multirow{2}[2]{*}{MCL} & \multicolumn{2}{l|}{DGI} & 62.44\tiny{±10.12} & 60.75\tiny{±10.97} & 58.15\tiny{±9.63} & 58.95\tiny{±9.60} & 70.22\tiny{±11.43} & 70.12\tiny{±12.46} & 60.83\tiny{±10.84} & 62.06\tiny{±10.16} \bigstrut[t]\\
              & \multicolumn{2}{l|}{InfoG} & 62.87\tiny{±9.52} & 62.37\tiny{±9.67} & 60.88\tiny{±9.97} & 60.44\tiny{±9.61} & 72.46\tiny{±8.71} & 72.94\tiny{±8.68} & 61.75\tiny{±9.76} & 61.37\tiny{±9.85} \bigstrut[b]\\
    \hline
    \multirow{2}[2]{*}{EGS} & \multicolumn{2}{l|}{GCC} & \underline{63.45\tiny{±9.82}} & 62.39\tiny{±9.08} & 60.44\tiny{±9.54} & 60.29\tiny{±10.33} & 70.97\tiny{±10.31} & 72.48\tiny{±11.36} & 61.27\tiny{±9.66} & 61.38\tiny{±10.72} \bigstrut[t]\\
              & \multicolumn{2}{l|}{EGI} & 63.38\tiny{±8.93} & \underline{63.58\tiny{±8.02}} & \underline{61.82\tiny{±8.53}} & 61.57\tiny{±8.27} & \underline{73.46\tiny{±8.49}} & \underline{73.28\tiny{±8.68}} & 60.89\tiny{±9.87} & \underline{62.41\tiny{±8.50}} \bigstrut[b]\\
    \hline
    \multirow{3}[2]{*}{Ours} & \multicolumn{2}{l|}{STP} & 53.92\tiny{±12.82}* & 54.61\tiny{±11.76}* & 55.51\tiny{±15.74}* & 56.73\tiny{±16.23}* & 61.18\tiny{±14.57}* & 62.88\tiny{±15.58}* & 55.29\tiny{±12.38}* & 57.31\tiny{±14.72}* \bigstrut[t]\\
    & \multicolumn{2}{l|}{MTP} & 60.37\tiny{±12.42}* & 61.64\tiny{±11.83}* & 59.41\tiny{±11.62}* & 59.92\tiny{±13.37}* & 67.65\tiny{±12.26}* & 68.38\tiny{±12.94}* & 60.54\tiny{±13.83}* & 59.46\tiny{±12.33}* \bigstrut[t]\\ 
    & \multicolumn{2}{l|}{\ours} & \textbf{68.84\tiny{±8.26}}* & \textbf{68.45\tiny{±8.96}}* & \textbf{66.57\tiny{±7.67}}* & \textbf{68.31\tiny{±9.39}}* & \textbf{77.80\tiny{±9.76}}* & \textbf{77.22\tiny{±8.74}}* & \textbf{67.51\tiny{±8.67}}* & \textbf{67.74\tiny{±8.59}}* \bigstrut[b]\\
    \hline
    \end{tabular}
  \label{tab:compare}}
\end{table*}

We evaluate the effectiveness of \ours through extensive experiments on real brain network datasets, with a focus on the following research questions: 
\begin{itemize}[nosep,leftmargin=*]
\item \textbf{RQ1}: How does \ours compare with other unsupervised GNN pre-training frameworks adapted to the scenario of brain networks?
\item \textbf{RQ2}: What is the contribution of each major component in \ours to the overall performance?
\item \textbf{RQ3}: How does the choice of sampling method affect model convergence and performance?
\item \textbf{RQ4}: How effective is the variance-based sorting in aligning virtual ROIs among different parcellation systems?
\end{itemize}

\paragraph{Datasets, Configurations, and Metrics.} 
Our experiments are conducted on three real-world brain network datasets: PPMI, BP, and HIV. The PPMI dataset is parcellated using the Desikan-Killiany84 atlas template and includes brain networks from 718 subjects, 569 of whom are Parkinson’s Disease (PD) patients and 149 are Healthy Control (HC). The networks are constructed using three tractography algorithms: Probabilistic Index of Connectivity (PICo), Hough voting (Hough), and FSL. The BP dataset is parcellated using the Brodmann82 template and includes resting-state fMRI and DTI modalities from 97 subjects, 52 of whom have Bipolar I disorder and 45 are HCs. The HIV dataset is parcellated using the AAL90 template and includes fMRI and DTI modalities from 70 subjects, with 35 early HIV patients and 35 HCs. We pre-train the model on the PPMI dataset and evaluate the downstream performance on BP and HIV. Further details about the datasets can be found in Appendix \ref{appendix:datasets}.

\ours employs GCN as the backbone for the GNN \citep{kipf2016semi} encoder. We also benchmark \ours with GAT \citep{velivckovic2018graph} and GIN \citep{xu2019powerful}, and the results are provided in Appendix \ref{sec:gatgin}.
The hyperparameter settings are described in detail in Appendix \ref{appendix:hyperparameters}. The hyperparameter tuning follows the standard designs in related studies such as in \citep{yang2021autism, wein2021graph, hu2021gat}. The downstream evaluation is binary graph classification for disease prediction.
To assess the performance, we use the two widely used metrics in the medical field \citep{li2021braingnn, cui2022braingb}: accuracy score (ACC) and the area under the receiver operating characteristic curve (AUC).

\subsection{Overall Performance Comparison (RQ1)}
\label{sec:RQ1}
We present a comprehensive comparison of the target performance between the proposed \ours and popular unsupervised learning strategies in Table \ref{tab:compare}.
To fairly compare the methods, we apply atlas mapping pre-processing and the multi-dataset learning backbone discussed in section \ref{sec:multi} to all methods. The purpose of this comparison is to effectively highlight the impact of the proposed two-level contrastive pre-training and we will further analyze the effect of atlas mapping in subsequent subsections.
In addition, for a clearer presentation, we group the selected baselines according to their optimization strategies:
\begin{itemize}[nosep,leftmargin=*]
\item No pre-training (NPT): the backbone with randomly initialized parameters for target evaluation.
\item Non-CL-based (NCL): methods with cost functions regularized by co-occurrence agreement or link reconstruction, including Node2Vec \citep{grover2016node2vec}, DeepWalk \citep{perozzi2014deepwalk}, and VGAE \citep{Kipf2016VariationalGA}.
\item Single-scale CL (SCL): methods utilizing either node- or graph-level representations in the CL optimization, including GBT \citep{bielak2022graph}, ProGCL \citep{xia2022progcl}, and GraphCL \citep{you2020graph}.
\item Multi-scale CL (MCL): methods whose CL optimization utilizes both nodes- and graph-level representations, including DGI \citep{velickovic2019deep} and InfoG \citep{sun2019infograph}. 
\item Ego-graph sampling (EGS): methods whose contrastive samplings consider $k$-hop ego-networks as discriminative instances, which are the most similar to the proposed \ours, including GCC \citep{qiu2020gcc} and EGI \citep{zhu2021transfer}.
\item Our proposed two-level contrastive optimization (Ours): methods include single task pre-training (STP) in which we select the PICo modality of the PPMI study to be the only source task; multi-task pre-trainig (MTP) which does not utilize the MAML technique; and the full implementation of the \ours framework.
\end{itemize}

The experiments reveal the following insights:
\begin{itemize}[nosep,leftmargin=*]
    \item The proposed \ours consistently outperforms all the baselines, achieving a relative improvement of 7.34\%-13.30\% over the best-performing baselines and 31.80\%-38.26\% over the NPT setting. The results of \ours have been statistically compared against baselines using paired $t$-tests.  With a significance level set to 0.05, the largest two-tailed $p$ value is reported at 0.042, indicating that \ours demonstrates a statistically significant performance increase over other selected methods.
    \item Compared with the transductive methods of Node2Vec and DeepWalk, the GNN pre-trained by VGAE learns structure-preserving representations and achieves the best results in the NCL-type methods. This indicates the potential benefit of the locality-preserving regularizer design in \ours.
    \item Maximizing mutual information between augmented instances may hinder GNNs from learning a shared understanding of the entire dataset. For baselines belonging to the categories of SCL, MCL, and EGS, pre-training with non-augmented CL (InfoG, EGI) generally results in a 4.36\% relative improvement across both metrics and a 7.63\% relative decrease in performance variance compared to their augmentation-based counterparts (GBT, GraphCL, ProGCL, DGI, GCC). This explains why \ours does not employ data augmentation.
    \item Multi-scale MI promotes the capture of effective local ($\textit{i.e.,}$ node-level) representations that can summarize the global ($\textit{i.e.,}$ graph-level) information of the entire network. The MCL-type methods typically outperform the SCL-type ones by a relative gain of 2.68\% in ACC and 3.27\% in AUC.
    \item The group of baselines considering $k$-hop neighborhoods (EGS) presents the strongest performance, indicating the importance of local neighborhoods in brain network analysis. The proposed \ours, which captures this aspect through both node- and graph-level CL, is the only one that comprehensively captures the local neighborhoods of nodes.
    \item Learning from multiple tasks (MTP) brings significant improvement over STP, reporting a relative increase of 8.47\% in accuracy and 6.90\% in AUC. Furthermore, the full \ours framework with MAML-styled training achieves a relative improvement of 11.29\% in accuracy, 14.75\% in AUC, and a reduced variance over MTP, demonstrating its advantages in enhancing model generalizability.
\end{itemize}

\begin{table}
  \centering
    \caption{The four variants of sampling strategies.}
    \vspace*{5pt}
    \begin{tabular}{|l|l|l|l|l|l|}
    \hline
         & $\mathbf{S_1}$  & $\mathbf{S_2}$  & $\mathbf{S_3}$  & $\mathbf{S_4}$ \bigstrut\\
    \hline
    Var. 1 & --  & --  & /  & /  \bigstrut\\
    \hline
    Var. 2 & +  & --  & /   & /  \bigstrut\\
    \hline
    Var. 3 & +  & --  & --  & /  \bigstrut\\
    \hline
    Var. 4 & +  & +  & --  & /  \bigstrut\\
    \hline
    \end{tabular}
  \label{tab:variants}
  \vspace*{-15pt}
\end{table}
\vspace*{-10pt}
\begin{figure*}[h]
    \centering
    \subfigure{\includegraphics[width=0.24\textwidth]{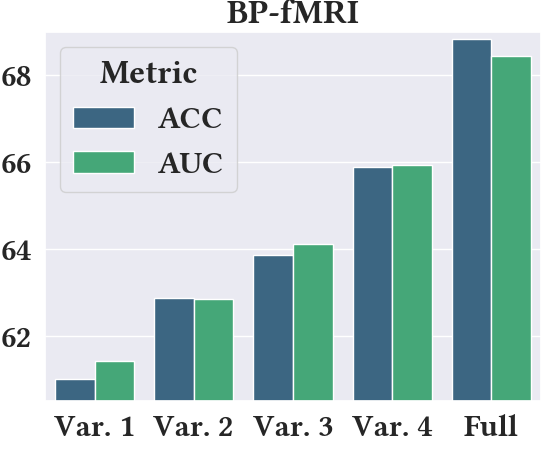}}
    \subfigure{\includegraphics[width=0.24\textwidth]{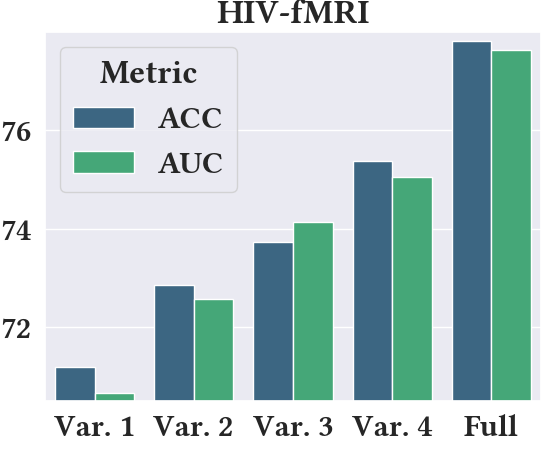}}
    \subfigure{\includegraphics[width=0.24\textwidth]{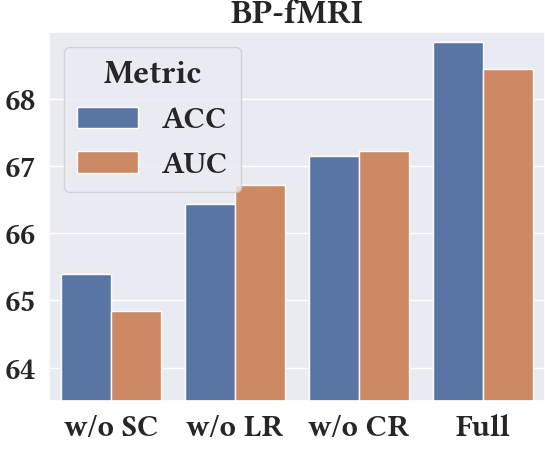}} 
    \subfigure{\includegraphics[width=0.24\textwidth]{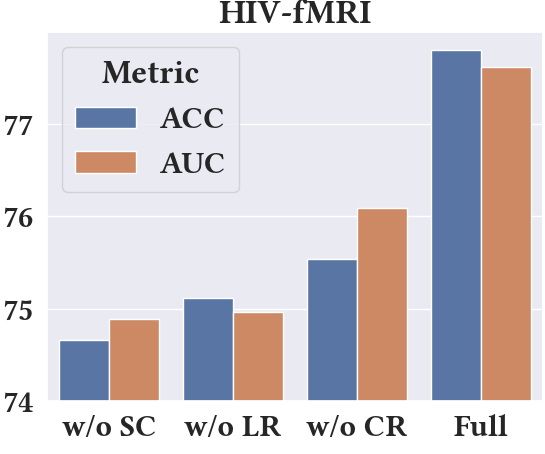}} 
    \vspace*{-5pt}
    \caption{Ablation comparisons on contrastive sampling choices (left two) and atlas mapping regularizers (right two). The $y$-axis refers to the numeric values of evaluated metrics (in \%). The setup of Var. 1 - 4 is described in Table \ref{tab:variants}. ``SC'', ``LR'', and ``CR'' are abbreviations for ``sparsity constraints'', ``locality regularizer'', and ``community (modularity-aware) regularizer'' respectively.}
    \vspace*{-10pt}
    \label{fig:RQ2}
\end{figure*}
\begin{figure*}
    \centering
    \subfigure[Pre-train loss on PPMI]{\includegraphics[width=0.3\textwidth]{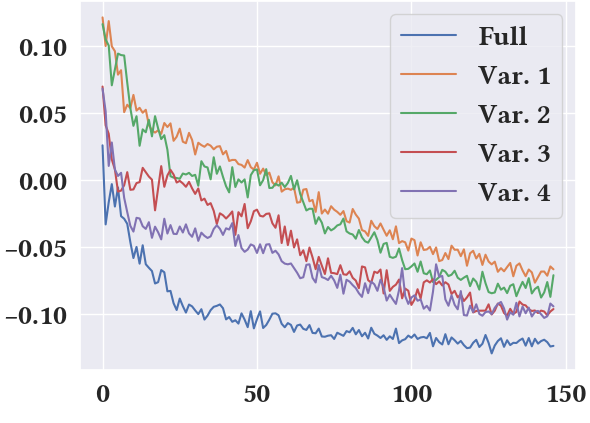} \label{fig: pretrain}} 
    \subfigure[Test ACC (\%) on HIV-fMRI]{\includegraphics[width=0.3\textwidth]{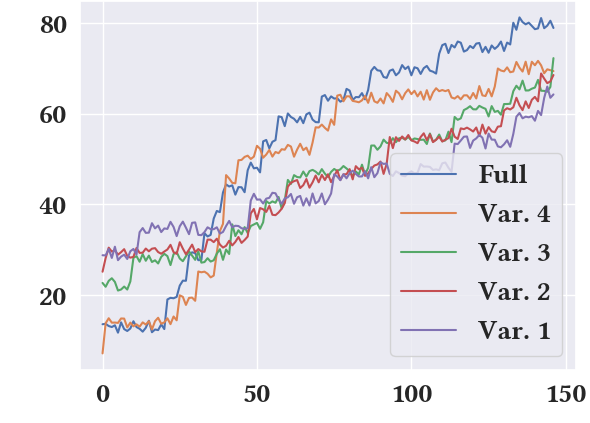} \label{fig: acc}} 
    \subfigure[Pre-train time (secs) on PPMI]{\includegraphics[width=0.3\textwidth]{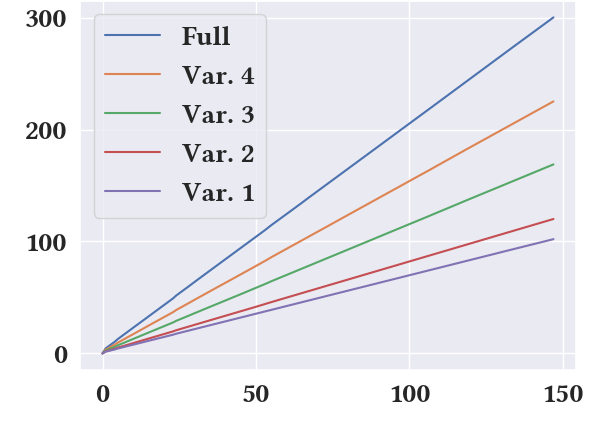} \label{fig: time}}
    \vspace*{-5pt}
    \caption{In-depth comparison among the four variants and the full model. The $x$-axis is epochs. Fig. (a) evaluates the trajectory of pre-training loss, Fig. (b) evaluates their respective testing accuracy on the fMRI view of the HIV dataset, and Fig. (c) reports the pre-training runtime in seconds.}
    \label{fig: RQ3}
\end{figure*}
\subsection{Ablation Studies (RQ2)}
\label{sec:RQ2}
We examine two key components of \ours - (1) the two-level contrastive sampling and (2) the atlas mapping regularizers. The best contrastive sampling configuration is fixed when examining the atlas regularizers, and all regularizers are equipped when examining the contrastive samplings.
The results, shown in Figure \ref{fig:RQ2} (with additional DTI version in Appendix \ref{sec:dti}), are analyzed based on the four possible variants of contrastive sampling listed in Table \ref{tab:variants}.
Our analyses yield the following observations: 
\textbf{(1)} leveraging $k$-hop neighborhood (\textit{i.e.,} positive $\mathbf{S_2}$) MI maximization brings visible performance gain, confirming its benefit in brain structure learning; \textbf{(2)} The extension to multi-graph CL (\textit{i.e.,} consideration of $\mathbf{S_3}$) facilitates the extraction of unique ROI knowledge, leading to improved results in Var.~3/4; \textbf{(3)} Var.~4 outperforms Var.~3 as it effectively summarizes of global (\textit{i.e.,} graph-level) information in local node representations;
\textbf{(4)} The full implementation of \ours brings a relative gain of 4.27\% in both metrics on top of Var.~4, highlighting the significance of considering shared substructure knowledge across multiple graphs (\emph{i.e.,} through the inclusion of $\mathbf{S_4}$).

The right-side sub-figures examine the impact of the atlas mapping regularizers by comparing the results of the full framework to those without the sparsity regularizer (w/o SR), the locality regularizer (w/o LR), and the community regularizer (w/o CR).
Two key observations are made: \textbf{(1)} The removal of SR leads to the greatest performance drop, emphasizing its crucial role in learning robust projections that can effectively handle noise and prevent over-fitting; 
\textbf{(2)} The inferior results when LR and CR are absent emphasize the importance of spatial sensitivity and blockwise feature information in brain network analysis. This supports our intuition to consider the relative positioning of ROIs in the 3D coordinate as well as knowledge on community belongings based on modularity measures.

\subsection{Analysis of Two-level Contrastive Sampling (RQ3)}
\label{sec:RQ3}
Figure \ref{fig: RQ3} offers insight into the pre-training convergence, target adaptation progression, and pre-training runtime consumption of the four sampling variants and the full framework. Key observations include: \textbf{(1)} As seen in Figure \ref{fig: pretrain}, all variants demonstrate efficient pre-training convergence due to the multi-dataset joint optimization inspired by MAML. The full model demonstrates the most optimal convergence, highlighting the advantage of learning shared neighborhood information in brain network data through two-level node contrastive sampling. \textbf{(2)} Figure \ref{fig: acc} shows the superiority of our design in terms of downstream adaptation performance compared to other variants.
\textbf{(3)} Figure \ref{fig: time} reveals that the more sophisticated the sampling considerations result in greater computational complexity for mutual information evaluation, leading to longer runtime for each pre-training epoch. However, the total time consumptions are all on the same scale.

\begin{figure*}
    \centering
    \subfigure[PPMI mapping]{\includegraphics[width=0.32\textwidth]{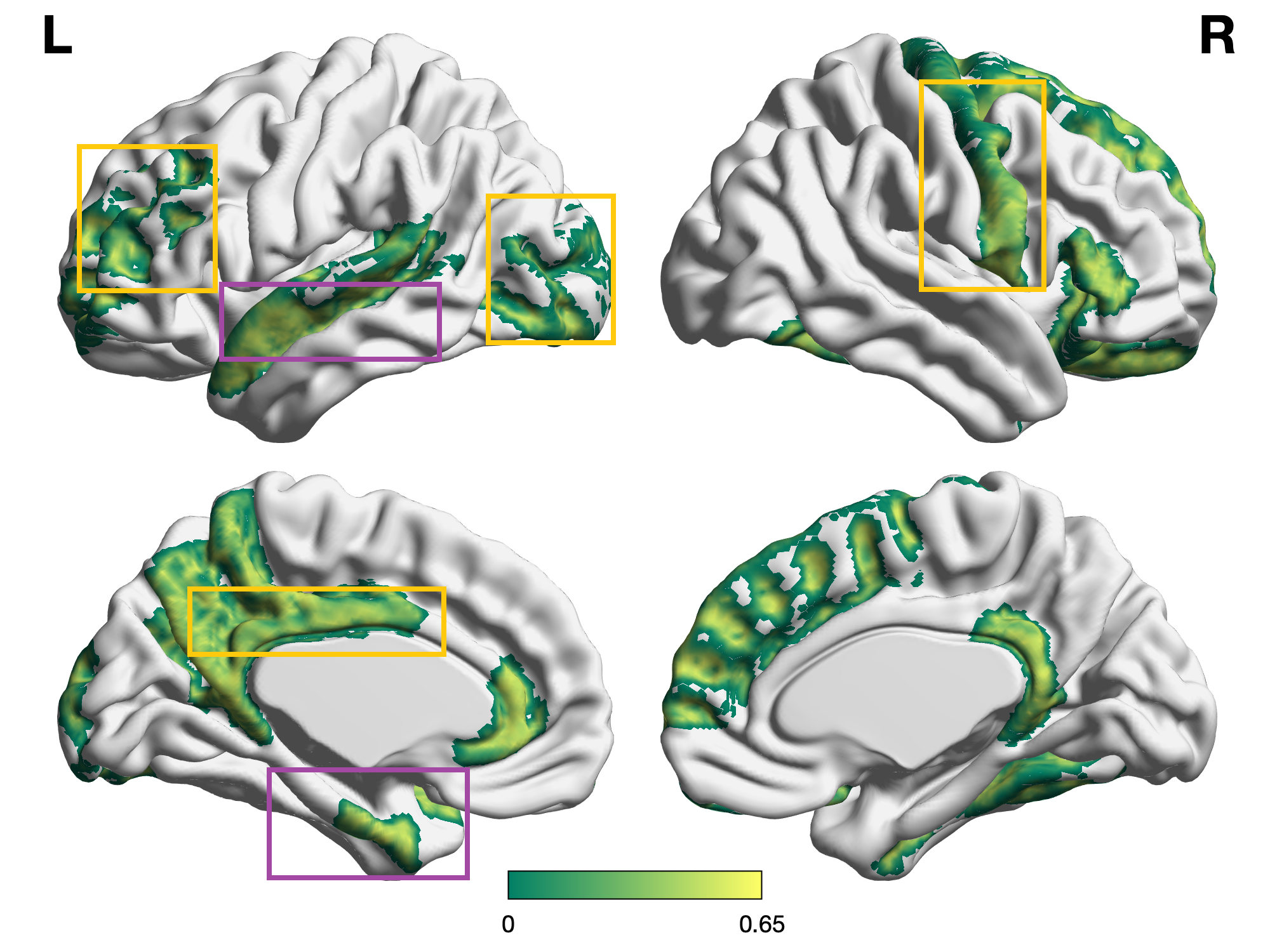}}
    \subfigure[BP mapping]{\includegraphics[width=0.32\textwidth]{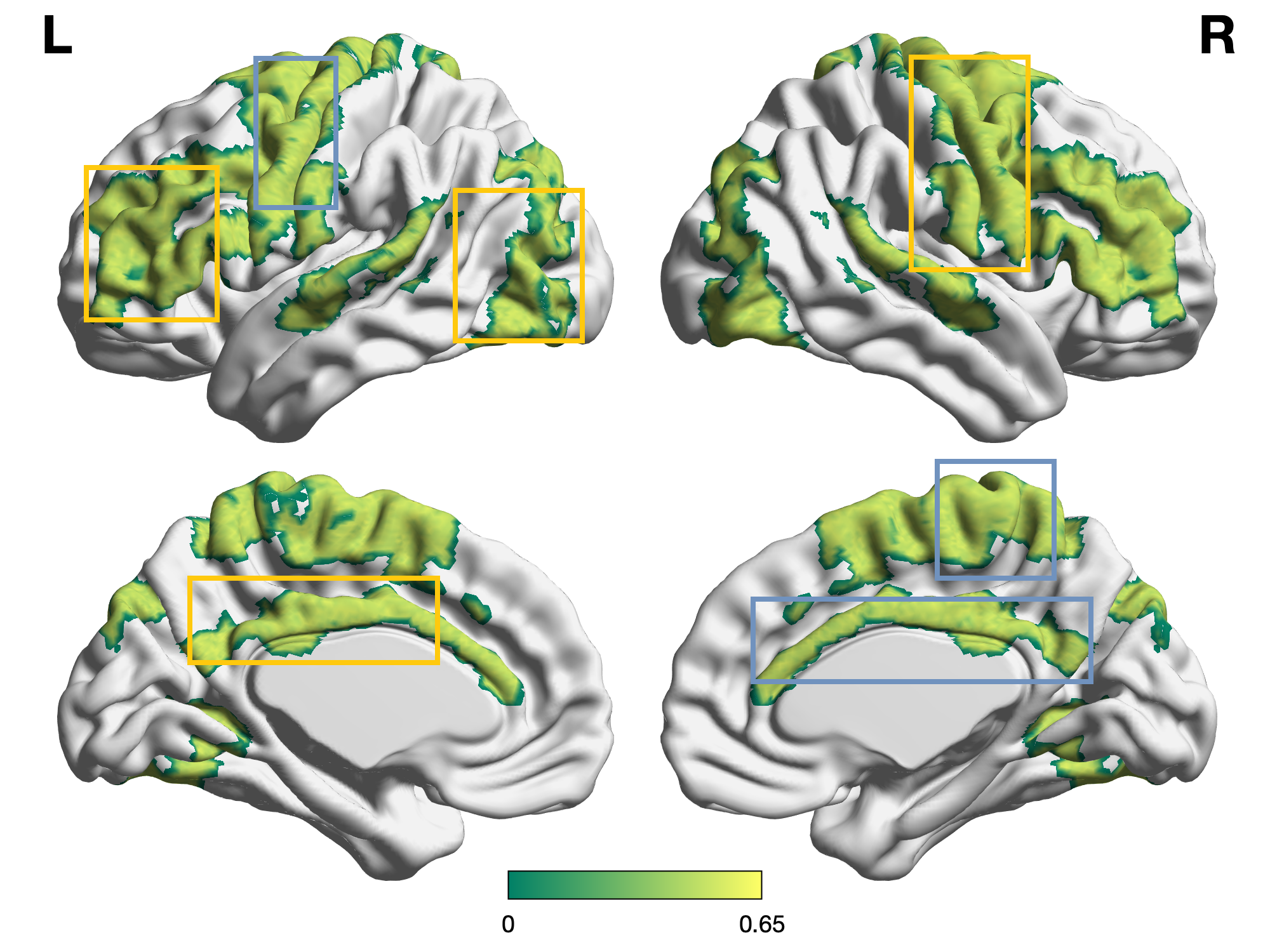}}
    \subfigure[HIV mapping]{\includegraphics[width=0.32\textwidth]{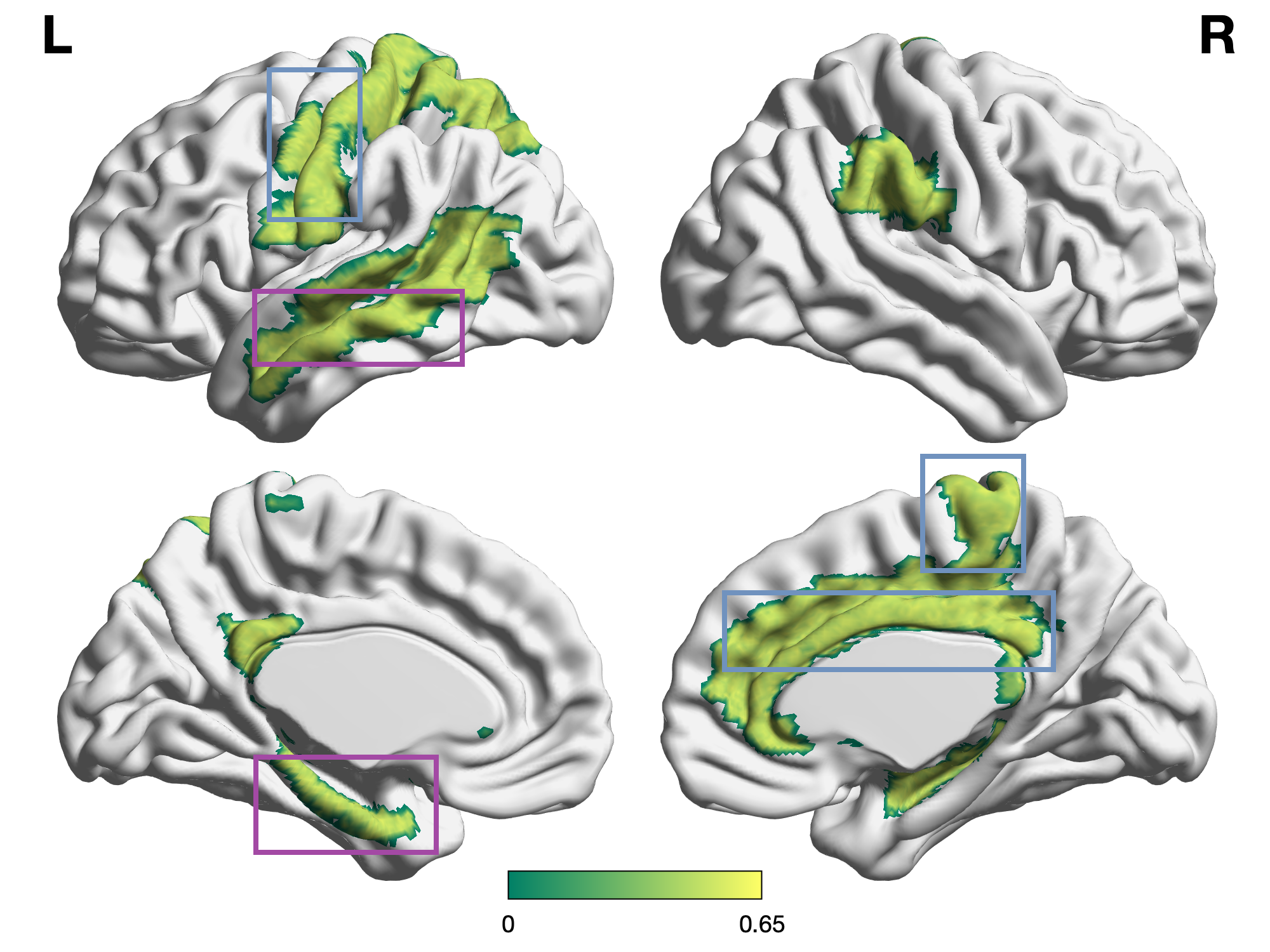}}
    \vspace*{-5pt}
    \caption{The virtual ROI mapping across the three investigated datasets. We highlight pairs of overlapping regions with colored boxes. In particular, we use gold boxes for the PPMI and BP mapping; blue boxes for the BP and HIV mapping; and purple boxes for the PPMI and HIV mapping.}
    \vspace*{-5pt}
    \label{fig: RQ4}
\end{figure*}
\subsection{Analysis of ROI Alignment (RQ4)}
\label{sec:RQ4}
To further validate the variance-based virtual ROI sorting, we select the top 2 virtual ROIs with the highest sample variances for each atlas template (\textit{i.e.,} dataset) and backtrack to locate their corresponding projected ROIs. The results are illustrated in Figure \ref{fig: RQ4}, which shows a 3D brain surface visualization highlighting the original ROIs. From this, we draw two main conclusions: 
$\textbf{(1)}$ There exists multiple regional overlaps between pairs of two atlas templates, reflecting some working effectiveness of our proposed solution as well as confirming the feasibility of converting between atlas templates.
$\textbf{(2)}$ It is relatively harder to find regions that overlap across all three atlas templates which shows a limitation of the proposed unsupervised ROI alignment scheme, suggesting a need to modify against the current variance-based heuristic which may inspire further study and research opportunity.

%% file: Sections/05_conclusions.tex
\section{Conclusion}
Brain network analysis for task-specific disease prediction has been a challenging task for conventional GNN frameworks due to the limited availability of labeled training data and the absence of a unifying brain atlas definition, which hinders efficient knowledge transfer across different datasets.
To address these challenges, we propose \ours, a novel unsupervised multi-dataset GNN pre-training that leverages a two-level node contrastive sampling to overcome data scarcity. Additionally, \ours incorporates atlas mapping through brain-network-oriented regularizers and variance-based sorting to address the issue of incompatible ROI parcellation systems in cross-dataset model adaptation in a data-driven way. 
Extensive experiments on real-world brain connectome datasets demonstrate the superiority and robustness of \ours in disease prediction and its clear advantage over various state-of-the-art baselines. As more brain network datasets become available, it will be intriguing to further validate its generalizability.

%% file: Sections/06_appendix.tex
\section{Autoencoder Structure Analysis}
\label{appendix:autoencoder}
\vspace*{-1mm}
\subsection{Bridging Reconstruction Minimization and Variance Maximization}
\label{sec:pcaequiv}
In this subsection, we briefly discuss how the reconstruction minimizing objective in one-layer AE can be cast to a variance-maximizing objective in PCA.
Assume given a data matrix $\mathbf{X} \in \mathbb{R}^{n\times d}$, its covariance matrix $\mathbf{\Sigma} = \mathbf{X}^\top \mathbf{X} \in \mathbb{R}^{n\times n}$, and a single-layer AE projection matrix $\mathbf{W} \in \mathbb{R}^{d\times m}$ with parameters randomly initialized from the continuous uniform distribution $\mathcal{U}(0, 1)$, the reconstruction objective is:
\begin{align*}
    \frac{1}{n} \lVert \mathbf{X} - \mathbf{X} \mathbf{W} \mathbf{W}^\top \rVert^2 &= \frac{1}{n} \text{tr} (  (\mathbf{X} - \mathbf{X} \mathbf{W} \mathbf{W}^\top )\\
    &\cdot (\mathbf{X} - \mathbf{X} \mathbf{W} \mathbf{W}^\top)^\top )\\
    &= \frac{1}{n} \text{tr} (  (\mathbf{X} - \mathbf{X} \mathbf{W} \mathbf{W}^\top) \\
    &\cdot (\mathbf{X}^\top - \mathbf{W} \mathbf{W}^\top \mathbf{X}^\top ) )\\
    &= \frac{1}{n} [ \text{tr}(\mathbf{X}\mathbf{X}^\top) - \text{tr}(\mathbf{X} \mathbf{W} \mathbf{W}^\top \mathbf{X}^\top)\\
    &- \text{tr}(\mathbf{X} \mathbf{W} \mathbf{W}^\top \mathbf{X}^\top) \\
    &+ \text{tr}(\mathbf{X}\mathbf{W}\mathbf{W}^\top\mathbf{W}\mathbf{W}^\top\mathbf{X}^\top) ]\\
    &= \frac{1}{n}[c_1 - 2\cdot \text{tr}(\mathbf{X}\mathbf{W}\mathbf{W}^\top\mathbf{X}^\top)\\
    &+ \text{tr}(\hat{\mathbf{X}}\hat{\mathbf{X}}^\top)]\\
    &= \frac{1}{n}[c_1 - 2\cdot \text{tr}(\mathbf{X}\mathbf{W}\mathbf{W}^\top\mathbf{X}^\top) + c_2]\\
    &= c_3 - c_4\cdot \text{tr}(\mathbf{W}^\top\mathbf{X}^\top\mathbf{X}\mathbf{W})\\
    &= c_3 - c_4\cdot \text{tr}(\mathbf{W}^\top\mathbf{\Sigma}\mathbf{W})
\end{align*}
Notice that $c_1, c_2, c_3, c_4$ are non-negative scalar constants that do not influence the overall optimization trajectory.
Hence, alternatively, the optimal AE projection also maximizes the sample variance $\text{tr}(\mathbf{W}^\top\mathbf{\Sigma}\mathbf{W})$, achieving an identical end goal of PCA transform.
Specifically, according to PCA, variance maximization is realized by constructing the projection $\mathbf{W}$ to contain the set of orthonormal eigenvectors of $\mathbf{\Sigma}$ that gives the largest eigenvalues \citep{hotelling1933analysis}.
That is, there is an orthogonality constraint on $\mathbf{W}$.
Minimizing the MSE reconstruction also results in an orthogonal $\mathbf{W}$:
\begin{equation*}
    \frac{1}{M}\lVert \mathbf{X} - \mathbf{X}\mathbf{W}\mathbf{W}^\top \rVert^2 = 0 \Rightarrow \mathbf{W}\mathbf{W}^\top = \mathbf{I}
\end{equation*}%
Therefore, the optimal AE projection $\mathbf{W}$ is also capturing a set of variance-maximizing orthogonal vectors.
Note that the AE optimized $\mathbf{W}$ is theoretically equivalent to the eigendecomposition of $\mathbf{\Sigma}$ if and only if the reconstruction loss is 0.
Therefore, in practice, the AE is, at best, an approximate solution to variance maximization. 

\subsection{Variance-based Sorting Procedure}
\label{sec:varsorting}
Following the discussion in \ref{sec:pcaequiv}, assuming a perfect optimization, the linear one-layer AE behaves similarly to PCA, and there is an equivalence relation between their respective objective functions.
Notice that in PCA, the eigenvalue of the covariance matrix $\mathbf{\Sigma}$ signifies the intensity of data variation along the direction of its corresponding eigenvector, which is essentially a column entry of the transformation matrix.
Then intuitively, given an optimized AE projection $\mathbf{W}$, we can examine, for each column of $\mathbf{W}$, its representativeness (\textit{i.e.,} data variance) of the data covariance with a scalar estimate (\textit{i.e.,} an eigenvalue-like scoring).
Inspired by the properties of eigendecomposition, we can approximate these estimates by measuring the distance of $\mathbf{W}$ w.r.t to the product of linearly transforming $\mathbf{W}$ through $\mathbf{\Sigma}$ by a scaling factor of $\bm{\lambda}$.
More specifically, we want to solve for $\bm{\lambda}$ such that $\mathbf{\Sigma} \bm{w} = \bm{\lambda} \bm{w}$ for every column vector $\bm{w} \in \mathbf{W}$. 
Under the PCA perspective, $\bm{\lambda}$ contains the variance estimate for each column-wise individual projection of $\mathbf{W}$. 
To this end, we detail the sorting procedure in Algorithm \ref{alg:variance}.\\

\begin{algorithm}
\caption{Overview procedure for variance-based sorting}\label{alg:variance}
\begin{algorithmic}[1]
\Input Original feature matrix $\mathbf{X} \in \mathbb{R}^{M \times M}$; AE optimized projection matrix $\mathbf{W} \in \mathbb{R}^{M \times D}$
\Initialize Scalar vector $\bm{\lambda} \in \mathbb{R}^D$; Small positive float $\epsilon$
\Output Sorted AE projection matrix $\Tilde{\mathbf{W}}$
\State Normalize the feature matrix: $\mathbf{X_n} \gets \mathbf{X} / \lVert\mathbf{X} \rVert$
\State Compute data covariance matrix: $\mathbf{\Sigma} \gets \mathbf{X_n}^\top \mathbf{X_n}$
\State Solve for $\bm{\lambda}$ such that $| \mathbf{\Sigma} \mathbf{W} - \mathbf{W} \odot \text{diag}(\bm{\lambda}) | \leqslant \epsilon$ 
\State Sort column vectors $\bm{w} \in \mathbf{W}$  according to (sorted) decreasing order of $\bm{\lambda}$ to obtain $\Tilde{\mathbf{W}}$
\end{algorithmic}
\end{algorithm}

\section{Dataset Details}
\label{appendix:datasets}
\begin{itemize}[nosep,leftmargin=*]
    \item \textbf{Parkinson’s Progression Markers Initiative (PPMI)}: We pre-train the model on large-scale real-life Parkinsons Progression Markers Initiative (PPMI) data of 718 subjects, where 569 subjects are Parkinson’s Disease (PD) patients and the rest 149 are Healthy Control (HC) ones. Eddy-current and head motion correction are performed using FSL\footnote{\url{https://fsl.fmrib.ox.ac.uk/fsl/fslwiki/}} and the brain networks are extracted using the same tool. The EPI-induced susceptibility artifacts correction is handled using Advanced Normalization Tools (ANT)\footnote{\url{http://stnava.github.io/ANTs/}}. In the meantime, 84 ROIs are parcellated from T1-weighted structural MRI using Freesurfer\footnote{\url{https://surfer.nmr.mgh.harvard.edu/}\label{freesurfer}}. The brain networks are constructed using three whole brain tractography algorithms namely the Probabilistic Index of Connectivity (PICo), Hough voting (Hough), and FSL. Each resulted network for each subject is 84 $\times$ 84. Each brain network is normalized by the maximum value to avoid computation bias for the later feature extraction and evaluation, since matrices derived from different tractography algorithms differ in scales and ranges. 
    \item \textbf{Bipolar Disorders (BP)}: This local dataset is composed of the resting-state fMRI and DTI image data of 52 Bipolar I subjects who are in euthymia and 45 Healthy Controls (HCs) with matched age and gender \citep{cao2015identification, ma2017multi}. The fMRI data was acquired on a 3T Siemens Trio scanner using a {T2}$^*$ echo planar imaging (EPI) gradient-echo pulse sequence with integrated parallel acquisition technique (IPAT) and DTI data were acquired on a Siemens 3T Trio scanner. The brain networks are constructed using the CONN\footnote{\url{http://www.nitrc.org/projects/conn/}} toolbox. We performed the normalization and smoothing after first realigning and co-registering the raw EPI pictures. After that, the signal was regressed to remove the confounding effects of the motion artifact, white matter, and CSF. The 82 cortical and subcortical gray matter regions produced by Freesurfer were identified, and pairwise signal correlations were used to build the brain networks.
    \item \textbf{Human Immunodeficiency Virus Infection (HIV)}: This local dataset involves fMRI and DTI brain networks for 70 subjects, with 35 of them early HIV patients and the other 35 Healthy Controls (HCs). These two groups of subjects do not differ in demographic distributions such as age and biological sex. The preprocessings for fMRI including brain extraction, slice timing correction and realignment are managed with the DPARSF\footnote{\url{http://rfmri.org/DPARSF/}} toolbox, while the preprocessings for DTI such as distortion correction are finished with the help of FSL toolbox. Finally, brain networks with 90 regions of interest are constructed based on the automated anatomical labeling (AAL) \citep{tzourio2002automated}.
\end{itemize}
\begin{table*}
  \centering
  \caption{Disease prediction performance of our framework using GAT and GIN. The best performer is highlighted in bold. }
  \resizebox{0.95\textwidth}{!}{
    \begin{tabular}{lr|llllllll}
    \hline
    \multicolumn{2}{l}{\multirow{2}[4]{*}{Method}} & \multicolumn{2}{c}{BP-fMRI} & \multicolumn{2}{c}{BP-DTI} & \multicolumn{2}{c}{HIV-fMRI} & \multicolumn{2}{c}{HIV-DTI} \bigstrut\\
\cline{3-10}    \multicolumn{2}{l}{} & \multicolumn{1}{c}{ACC} & \multicolumn{1}{c}{AUC} & \multicolumn{1}{c}{ACC} & \multicolumn{1}{c}{AUC} & \multicolumn{1}{c}{ACC} & \multicolumn{1}{c}{AUC} & \multicolumn{1}{c}{ACC} & \multicolumn{1}{c}{AUC} \bigstrut\\
    \hline
    \multicolumn{2}{l|}{Ours w/ GCN} & \textbf{68.84\tiny{±8.26}} & 68.45\tiny{±8.96} & \textbf{66.57\tiny{±7.67}} & \textbf{68.31\tiny{±9.39}} & \textbf{77.80\tiny{±9.76}} & 77.22\tiny{±8.74} & \textbf{67.51\tiny{±8.67}} & \textbf{67.74\tiny{±8.59}} \bigstrut[t]\\
    \multicolumn{2}{l|}{Ours w/ GAT} & 66.96\tiny{±9.71} & \textbf{69.68\tiny{±9.61}} & 64.23\tiny{±10.47} & 63.76\tiny{±10.49} & 74.93\tiny{±10.35} & 75.78\tiny{±11.12} & 65.84\tiny{±9.74} & 66.51\tiny{±12.07} \\
    \multicolumn{2}{l|}{Ours w/ GIN} & 66.30\tiny{±8.77} & 68.92\tiny{±9.37} & 64.48\tiny{±9.83} & 66.44\tiny{±8.58} & 75.96\tiny{±9.56} & \textbf{77.63\tiny{±10.10}} & 67.36\tiny{±9.26} & 65.95\tiny{±11.76} \\
    \hline
    \end{tabular}}
  \label{tab:gatgin}
\end{table*}
\begin{figure*}
    \centering
    \subfigure{\includegraphics[width=0.24\textwidth]{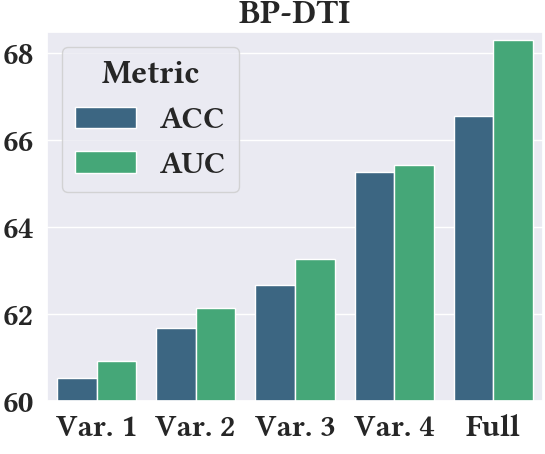}}
    \subfigure{\includegraphics[width=0.24\textwidth]{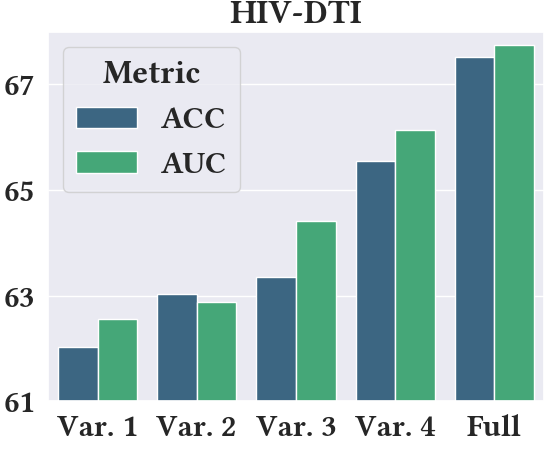}}
    \subfigure{\includegraphics[width=0.24\textwidth]{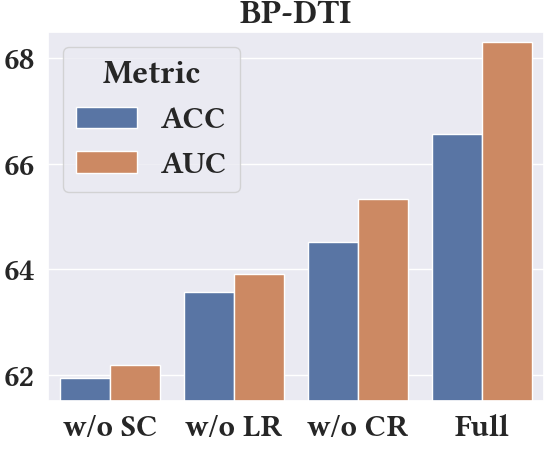}} 
    \subfigure{\includegraphics[width=0.24\textwidth]{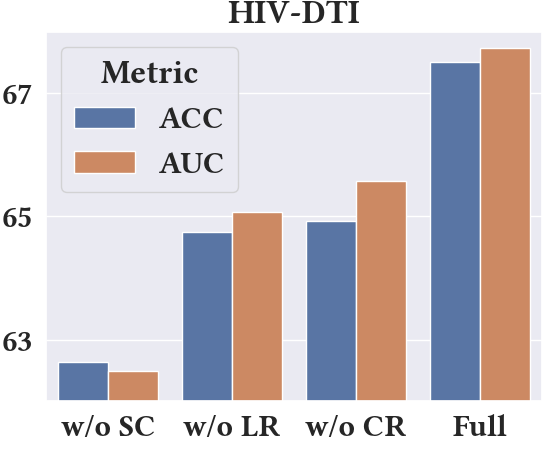}} 
    \vspace*{-2mm}
    \caption{Additional ablation comparisons on DTI views. The left two subfigures refer to contrastive sampling considerations and the right two subfigures refer to atlas mapping regularizers. The $y$-axis refers to the numeric values of evaluated metrics (in \%). We benchmark our results on the DTI modality of the BP and HIV dataset in this Appendix.}
    \vspace*{-1mm}
    \label{fig:additionalablation}
\end{figure*}
\section{Hyperparameter Setting}
\label{appendix:hyperparameters}
\paragraph{GNN Setup.}
The GCN encoder is composed of 4 graph convolution layers with hidden dimensions of 32, 16, 16, and 8.
Similarly, the GAT encoder is built from 4 graph attention layers with hidden dimensions of 32, 16, 16, and 8.
Regarding GIN, which is slightly different, the encoder consists of 4 MLP layers with each MLP containing 2 linear layers with a unifying hidden dimension of 8.
\paragraph{Pre-training Pipeline Setup.}
For two-level node contrastive sampling, we set $k=2$ as the radius regarding $k$-hop neighborhood sampling for $\mathbf{S_1}$ and $\mathbf{S_4}$.
To enable efficient computation on multi-graph MI evaluation, we resort to mini-batching and we set a default batch size of 32.
In addition, we leverage the popular Adam \citep{DBLP:journals/corr/KingmaB14} optimizer with the learning rate set to 0.002 as well as the cosine annealing scheduler \citep{DBLP:conf/iclr/LoshchilovH17} to facilitate GNN training.
In general, a complete pre-training cycle takes 400 epochs with an active deployment of early stopping.
\paragraph{Atlas Mapping Regularizer Setup.}
Following the discussion in section \ref{sec:regularizers}, the total running loss of the AE projection is given as:
\begin{equation}
    \mathcal{L} = \mathcal{L}_{\text{rec}} + \alpha \mathcal{L}_{\text{loc}} + \beta \mathcal{L}_{\text{com}} + \gamma \mathcal{L}_{\text{KL}},
\end{equation}%
in particular, we set $\alpha,\, \beta = 0.8$ and $\gamma = 0.01$.
The one-layer AE encoder transforms the feature signals from all given datasets into a universally projected dimension of 32.
For the details of locality-preserving regularizer (\textit{i.e.,} $\mathcal{L}_{\text{loc}}$), the transition matrix $\mathbf{T}$ is built from the 5-nearest-neighbor graph from the 3D coordinates of each atlas templates.
For the sparsity-oriented regularizer (\textit{i.e.,} $\mathcal{L}_{\text{KL}}$), the target sparsity value $\rho$ is set to $1e^{-5}$.
The overall optimization process, which is similar to model pre-training, takes a total of 100 epochs with a learning rate of 0.02.
\paragraph{Downstream Evaluation Setup.}
For each target evaluation, the fine-tuning process features a 5-fold cross-validation, which approximately splits the dataset into 70\% training, 10\% validation, and 20\% testing.
To prevent model over-fitting, we implement a $L_2$ penalty with a coefficient of $1e^{-4}$. Overall, the model fine-tuning process, which is nearly identical to the other two training procedures, takes a total of 200 epochs with a learning rate of 0.001 and a cosine annealing scheduler.
\balance
\section{Additional Experiment}
\subsection{Performance with GAT and GIN}
\label{sec:gatgin}

Table \ref{tab:gatgin} reports the downstream performance of our full framework using GAT and GIN as backbone encoders. In general, the two encoders deliver inferior performance compared to GCN, which suggests that complex GNN convolutions (\textit{e.g.,} GAT and GIN) might not be as effective as they seem when learning on brain network datasets.

\subsection{Additional Ablation Studies on DTI}
\label{sec:dti}
Figure \ref{fig:additionalablation} presents our ablation studies on the DTI view following the same setup as discussed in Section \ref{sec:RQ2}.
We draw similar conclusions from the DTI-based analysis where each constituent component of our two-level sampling consideration as well as the atlas mapping mechanism has proven positive contribution and significance towards the overall performance and robustness.